\documentclass[10pt,reqno]{amsart}

\usepackage{amsmath,amssymb,amsfonts,dsfont}
\usepackage{cite,graphicx}

\PassOptionsToPackage{linktocpage}{hyperref}

\renewcommand{\phi}{\varphi}
\newcommand{\cv}{\mathcal{V}}
\newcommand{\trans}{{\scriptscriptstyle \mathrm{T}}}
\newcommand{\LPT}{{L_2(\mathds{T})}}
\newcommand{\LPTT}{{L_2(\mathds{T}^2)}}
\newcommand{\BasisT}{\{q(i,t)\}_{i=0}^\infty}
\newcommand{\BasisTT}{\{q(i,t) \otimes q(j,\tau)\}_{i,j=0}^\infty}

\author{Konstantin A. Rybakov}

\title[On Spectral Approach to the Synthesis of Shaping Filters]{On Spectral Approach to the Synthesis of Shaping Filters\footnote{\MakeLowercase{rkoffice@mail.ru}}}

\begin{document}

\maketitle

\textbf{Abstract.} This paper describes various approaches to modeling a random process with a given rational power spectral density. The main attention is paid to the spectral form of mathematical description, which allows one to obtain a relation for the shaping filter using a transfer function without any additional calculations. The paper provides all necessary relations for the implementation of the shaping filter based on the spectral form of mathematical description.

\vskip 0.5ex


\vskip 0.5ex


\section{Introduction}\label{secIntro}

The paper considers various approaches to modeling a random process with a given power spectral density that is specified as a rational function, and the distribution of this random process is assumed to be Gaussian. Thus, we can speak about the problem of synthesizing a shaping filter, i.e., a linear continuous system whose input signal is the Gaussian white noise and whose output signal is a random process with desired probability characteristics. The description of shaping filters is traditionally based on power spectral densities or stochastic differential equations~\cite{Ste_17}.

The problem of synthesizing a shaping filter is relevant and in demand in a number of applications. We can emphasize problems of information processing in navigation\cite{Ste_17}, problems of radio engineering~\cite{TihHar_91}, and many others. It is needed for statistical modeling of complex engineering systems subject to random effects, e.g., in problems of aircraft or rotorcraft flight dynamics in turbulence~\cite{LiuAbaCapManFu_Aero2022, JiLuChen_Math2022}. The last example, namely the Dryden turbulence model, is considered in the recent paper~\cite{KhruRum_AT24}, which proposes a new method to obtain a system of stochastic differential equations for a given power spectral density (strictly speaking, for a transfer function that determines the power spectral density), i.e., one of the methods for synthesizing a shaping filter. There are also other methods to obtain a system of stochastic differential equations~\cite{PugSin_90}. However, it is interesting to note that despite the availability of such methods with a simple implementation, the standard~\cite{MIL-STD-1797A} recommends for modeling the turbulence to use the numerical scheme corresponding to first-order stochastic differential equations, although only two of six power spectral densities in the Dryden turbulence model correspond to first-order equations. The remaining power spectral densities correspond to second- and third-order equations.

The shaping filter can be constructively defined not only by stochastic differential equations. In general, for linear continuous systems with deterministic or random input signals, various forms of mathematical description can be used, on the basis of which methods and algorithms for solving analysis, synthesis, and identification problems are proposed. Four forms are distinguished~\cite{SolSemPeshNed_79, PanBor_16}:

(1) description by linear differential equations;

(2) description by response functions;

(3) description by integral transforms;

(4) description by spectral transforms.

The description of linear discrete systems assumes similar forms, of course, taking into account the discreteness of time, e.g., linear difference equations are used instead of linear differential equations.

The aim of this paper is to complement the paper~\cite{KhruRum_AT24} and to give a broader description of methods for modeling a random process with a given power spectral density, as well as their relation in the context of different forms of mathematical description for linear continuous systems. The main attention is paid to the spectral form of mathematical description, which allows one to obtain a relation for the shaping filter using a transfer function without any additional calculations. And this relation forms the basis for approximate modeling of a random process with desired probability characteristics. This approach allows one to obtain an approximation of a random process in continuous time and to calculate the mean square approximation error exactly. The paper contains an addition to~\cite{KhruRum_AT24} in terms of obtaining a system of stochastic differential equations that defines the shaping filter.

The rest of this paper is organized as follows. Section~\ref{secDiff} discusses the description of the shaping filter by differential equations. Section~\ref{secIRF} defines the shaping filter by the impulse response function. A new approach for the description of the shaping filter using the spectral form is proposed in Section~\ref{secSpectral}, and corresponding examples are given in Section~\ref{secExamples}. The paper is summarized in Section~\ref{secConcl}. Appendix contains all necessary relations for the implementation of the shaping filter based on the spectral form of mathematical description.

\section{Description of the Shaping Filter by Differential Equations}\label{secDiff}

We assume that the power spectral density $S(\omega)$ of the centered stationary random process $x(t)$ is given; it is defined by the Fourier transform $\mathbb{F}$ of the covariance function $R(\eta)$, i.e.,
\[
  \mathrm{E} x(t) = 0, \ \ \ \mathrm{E} x(t) x(t+\eta) = R(\eta), \ \ \ S(\omega) = \mathbb{F} [R(\eta)] = \int_{-\infty}^{+\infty} R(\eta) \mathrm{e}^{-\mathrm{i} \omega \eta} d\eta,
\]
and this power spectral density is represented as
\begin{equation}\label{eqSHrel}
  S(\omega) = |H(\mathrm{i} \omega)|^2.
\end{equation}

In the above formulae, $t$ and $\eta$ denote the time, $\omega$ is the frequency, $\mathrm{i}$ is the imaginary unit, $H(s)$ is the rational function of a complex variable $s$:
\begin{equation}\label{eqHdef}
  H(s) = \frac{M(s)}{D(s)} = \frac{b_m s^m + \ldots + b_1 s + b_0}{a_n s^n + \ldots + a_1 s + a_0},
\end{equation}
$a_0,a_1,\dots,a_n$ and $b_0,b_1,\dots,b_m$ are real numbers, $a_n,b_m \neq 0$, $n > m$, and $\mathrm{E}$ means the mathematical expectation.

Then $x(t)$ is the output signal of a one-dimensional linear stationary system called the shaping filter~\cite{PugSin_90, Ste_17} whose input signal $g(t)$ is the white noise, $H(s)$ is the transfer function, and $H(\mathrm{i} \omega)$ is the frequency response function~\cite{PanBor_16}. The random process $x(t)$ is assumed to be Gaussian, i.e., $g(t)$ is the standard Gaussian white noise.

The transfer function $H(s)$ corresponds to the following differential equation~\cite{PanBor_16}:
\begin{equation}\label{eqSDE}
  a_n x^{(n)}(t) + \ldots + a_1 x'(t) + a_0 x(t) = b_m g^{(m)}(t) + \ldots + b_1 g'(t) + b_0 g(t).
\end{equation}

If we assume that $g(t)$ is a sufficiently smooth function, i.e., its derivatives up to order $m$ exist and are continuous, then the (ordinary) differential equation does not raise any questions, and its solution is strictly defined. But if $g(t)$ is the standard Gaussian white noise, then the (stochastic) differential equation can be understood only formally even for $m = 0$. Sometimes the correctness of typical white noise definitions is questioned~\cite{GihScoYad_88}; however, here not only the white noise is used but also its derivatives.

The main result of~\cite{KhruRum_AT24} is a new method to obtain a system of linear stochastic differential equations that do not contain white noise derivatives:
\begin{equation}\label{eqSysSDE}
  \bar x'(t) = A \bar x(t) + B g(t),
\end{equation}
where $\bar x(t)$ is the vector-valued random process of size $n^*$, $A$ is the square matrix of size $n^* \times n^*$, and $B$ is the column matrix of size $n^* \times 1$, $n^* \geqslant n$.

A strict meaning can be given to this equation if we rewrite it as the stochastic integral equation using one of the stochastic integral definitions:
\begin{equation}\label{eqSysSDEInt}
  \bar x(t) = \bar x_0 + \int_0^t A \bar x(\tau) d\tau + \int_0^t B dw(\tau) = \bar x_0 + A \int_0^t \bar x(\tau) d\tau + B w(\tau),
\end{equation}
where $\bar x_0$ is the initial random vector of size $n^*$, and $w(t)$ is the standard Wiener process. Since the white noise is additive, this equation can be understood in different ways: in the sense of It\^o, Stratonovich, or Ogawa~\cite{Oks_03, Oga_17}. The solutions to such equations are stochastically equivalent.

The random process $x(t)$ is expressed as follows:
\begin{equation}\label{eqSysSDEOut}
  x(t) = C \bar x(t),
\end{equation}
where $C$ is the row matrix of size $1 \times n^*$.

Equations~\eqref{eqSysSDE} and~\eqref{eqSysSDEOut} define a multidimensional linear stationary system; $\bar x(t)$ is its state, and $x(t)$ is the output signal.

The optimal variant is realized for $n^* = n$. Therefore, the method described in~\cite{PugSin_90} and mentioned in~\cite{KhruRum_AT24} seems more preferable for the transition from the equation~\eqref{eqSDE} to the equation~\eqref{eqSysSDE}. In the method from~\cite{PugSin_90}, the square matrix $A$ is defined as
\[
  A = \left[ \begin{array}{ccccc}
    0 & 1 & 0 & \cdots & 0 \\
    0 & 0 & 1 & \cdots & 0 \\
    \vdots & \vdots & \ddots & \ddots & \vdots \\
    0 & 0 & 0 & \cdots & 1 \\
    -a_0/a_n & -a_1/a_n & -a_2/a_n & \cdots & -a_{n-1}/a_n
  \end{array} \right],
\]
or
\[
  A_{ij} = \left\{ \begin{array}{ll}
    1 & \text{for} ~ i = j-1 \\
    \displaystyle -\frac{a_{j-1}}{a_n} & \text{for} ~ i = n \\
    0 & \text{otherwise},
  \end{array} \right.
  \ \ \ \ \ \ i,j = 1,2,\dots,n,
\]
i.e., it corresponds to the common transition from a linear differential equation to a normal system of linear differential equations. The determinant $\det(sE-A)$ differs from the characteristic polynomial $D(s)$ by the coefficient $a_n$ (or coincides with it for $a_n = 1$): $D(s) = a_n \det(sE-A)$, where $E$ is the identity matrix of size $n \times n$.

The remaining matrices from the right-hand side of the equation~\eqref{eqSysSDE} are given by expressions
\[
  B = [B_1 ~ B_2 ~ \cdots ~ B_n]^\trans, \ \ \ C = [1 ~ 0 ~ \cdots ~ 0]^\trans,
\]
where
\begin{align*}
  B_i = \left\{ \begin{array}{ll}
    0 & \text{for} ~ i = 1,2,\dots,n-m-1 \vphantom{\sum\limits_j^i} \\
    \displaystyle \frac{b_{n-i}}{a_n} & \text{for} ~ i = n-m \\
    \displaystyle \frac{1}{a_n} \biggl( b_{n-i} - \sum\limits_{j = n-m}^{i-1} a_{n-i+j} B_j \biggr) & \text{for} ~ i = n-m+1,\dots,n,
  \end{array} \right.
\end{align*}
i.e., $x(t) = \bar x_1(t)$.

For example, for the transfer function ($\alpha,\beta,\gamma,\delta$ are its parameters; $\alpha,\delta \neq 0$)
\[
  H(s) = \frac{\alpha(\beta s + 1)}{\delta s^2 + \gamma s + 1}
\]
and the corresponding second-order stochastic differential equation
\[
  \delta x''(t) + \gamma x'(t) + x(t) = \alpha \bigl( \beta g'(t) + g(t) \bigr),
\]
we have
\begin{align*}
  x(t) = \bar x_1(t), \ \ \ \bar x'_1(t) = \bar x_2(t) + \alpha \, \frac{\beta}{\delta} \, g(t), \ \ \
  \bar x'_2(t) = -\frac{1}{\delta} \, \bar x_1(t) - \frac{\gamma}{\delta} \, \bar x_2(t) + \alpha \, \frac{\delta - \beta \gamma}{\delta^2} \, g(t),
\end{align*}
i.e., we obtain the system of linear stochastic differential equations of the second order. The method proposed in~\cite{KhruRum_AT24} forms the system of linear stochastic differential equations of the sixth order for the same transfer function; the system itself is not given in~\cite{KhruRum_AT24} due to its cumbersomeness.

The method from~\cite{PugSin_90} has several advantages:

(1)\;the order of the system of stochastic differential equations coincides with the order of the original stochastic differential equation ($n^* = n$);

(2)\;the low computational complexity (a low number of algebraic operations to obtain the result);

(3)\;the generalization to linear non-stationary systems (coefficients of the left-hand and right-hand sides of the stochastic differential equation may depend on time).

Note that for linear non-stationary systems, the equation for elements of the matrix $A$ is preserved; it is only necessary to take into account that the left-hand and right-hand sides depend on time. The equation for elements of the column matrix $B$ is complicated~\cite{PugSin_90} but the row matrix $C$ remains unchanged.

Another method, considered in~\cite{Ver_13} and reviewed in~\cite{KhruRum_AT24}, is based on the obvious equality of both the transfer function $H(s)$ and the transfer function for the output signal of a multidimensional linear stationary system described by equations~\eqref{eqSysSDE} and~\eqref{eqSysSDEOut}, i.e.,
\begin{equation}\label{eqTFs}
  H(s) = C (sE-A)^{-1} B.
\end{equation}

It is advisable to choose both the square matrix $A$ and the row matrix $C$ as above, and it is recommended to find the column matrix $B$ from the equality condition for polynomials, namely numerators of rational functions in the left-hand and right-hand sides of the equality~\eqref{eqTFs}, of course, with the same denominators. For this, it is sufficient to equate coefficients at the same powers of a complex variable $s$.

However, it is possible to reduce calculations to algebraic operations with matrices, making them more convenient and oriented towards the use of computer algebra systems. The column matrix $B$ must be such that the equality~\eqref{eqTFs} holds for arbitrary $s$. So, we can choose $n$ different $s_k$, $k = 1,2,\dots,n$, for which $H(s_k)$ are finite, i.e., $s_k$ must not coincide with poles of the function $H(s)$ (this means that $D(s_k) \neq 0$). For them, the matrix $(s_k E - A)$ has the inverse, and the expression $C (s_k E - A)^{-1}$ defines the first row of the matrix $(s_k E - A)^{-1}$. Thus, we obtain the system of linear algebraic equations:
\[
  C (s_k E-A)^{-1} B = H(s_k), \ \ \ k = 1,2,\dots,n,
\]
or in the matrix form
\[
  Q B = H = [H(s_1) ~ H(s_2) ~ \dots ~ H(s_n)]^\trans,
\]
where $Q$ is the square matrix of size $n \times n$ whose $k$th row is given by $C (s_k E - A)^{-1}$. The solution to this system is $B = Q^{-1} H$.

The matrix $(sE-A)$ has a special form, namely
\[
  sE-A = \left[ \begin{array}{ccccc}
    s & -1 & 0 & \cdots & 0 \\
    0 & s & -1 & \cdots & 0 \\
    \vdots & \vdots & \ddots & \ddots & \vdots \\
    0 & 0 & 0 & \cdots & -1 \\
    a_0/a_n & a_1/a_n & a_2/a_n & \cdots & s + a_{n-1}/a_n
  \end{array} \right],
\]
or
\[
  (sE-A)_{ij} = \left\{ \begin{array}{ll}
    s & \text{for} ~ i = j, ~ j \neq n \\
    -1 & \text{for} ~ i = j-1 \displaystyle \vphantom{\frac{a_n}{a_n}} \\
    \displaystyle \frac{a_{n-1}}{a_n} & \text{for} ~ i = n, ~ j \neq n \\
    \displaystyle s + \frac{a_{j-1}}{a_n} & \text{for} ~ i = n, ~ j = n \\
    0 & \text{otherwise},
  \end{array} \right.
  \ \ \ \ \ \ i,j = 1,2,\dots,n,
\]
and the first row of the inverse matrix is formed by rational functions for which the general formula can be used:
\[
  \frac{M_j(s)}{D(s)}, \ \ \ M_j(s) = \sum\limits_{l = j}^n a_l s^{l-j},
\]
where $j$ is the column number, $j = 1,2,\dots,n$. Therefore, the column matrix $B$ satisfies the equation
\[
  V B = M = [M(s_1) ~ M(s_2) ~ \dots ~ M(s_n)]^\trans
\]
in which $V$ is the square matrix of size $n \times n$ with elements
\[
  V_{kj} = \sum\limits_{l = j}^n a_l s_k^{l-j}, \ \ \ k,j = 1,2,\dots,n.
\]
The solution to this equation is $B = V^{-1} M$.

Although the described method provides the same result as the method from~\cite{PugSin_90}, its computational complexity is higher (a large number of algebraic operations to obtain the result). It cannot be generalized to linear non-stationary systems.

Thus, the equation~\eqref{eqSDE} can be associated with the equation~\eqref{eqSysSDE}, or in the strict sense, with the equation~\eqref{eqSysSDEInt}. This implies that the form~\eqref{eqSDE} is admissible at the level of rigor that allows one to consider Langevin-type stochastic differential equations. However, for modeling the random process $x(t)$, the equation~\eqref{eqSysSDE} is preferable if one focuses only on the description of random processes by stochastic differential equations. Approximate modeling can be carried out using any suitable numerical method for solving stochastic differential equations~\cite{KloPla_92, Kuz_DUPU23, AveRyb_SJVM24}. An arbitrary numerical scheme provides a discrete (in discrete time) approximation of the random process $x(t)$.

\section{Description of the Shaping Filter Based on the Impulse Response Function}\label{secIRF}

In the previous section, we assume that a linear stationary system is defined by the transfer function~\eqref{eqHdef} from which the differential equation~\eqref{eqSDE} or the equivalent system of differential equations~\eqref{eqSysSDE} can be obtained. These equations relate input and output signals. But linear systems can be defined in other forms of mathematical description, in particular by an impulse response function.

By definition, the impulse response function $k(t,\tau)$ is the output signal of a linear system whose input signal is the Dirac delta function $\delta(t-\tau)$ with zero initial conditions. For linear stationary systems, it can be considered as a function of one variable, i.e., $k(t-\tau) = k(\eta)$, $\eta = t-\tau$. One way to find the impulse response function is shown below~\cite{PanBor_16}:
\begin{equation}\label{eqIRFfull}
  k(\eta) = b_m k_0^{(m)}(\eta) + \ldots + b_1 k'_0(\eta) + b_0 k_0(\eta),
\end{equation}
where $k_0(\eta)$ is the solution to the following initial value problem for the linear differential equation:
\begin{equation}\label{eqIRFshort}
  \begin{gathered}
    a_n k_0^{(n)}(\eta) + \ldots + a_1 k'_0(\eta) + a_0 k_0(\eta) = 0, \\
    k_0(0+) = k'_0(0+) = \ldots = k_0^{(n-2)}(0+) = 0, \ \ \ k_0^{(n-1)}(0+) = \frac{1}{a_n}.
  \end{gathered}
\end{equation}

The transfer function is the Laplace transform $\mathbb{L}$ of the impulse response function, and the corresponding frequency response function is defined by the Fourier transform $\mathbb{F}$ taking into account the physical realizability condition, i.e., $k(\eta) = 0$ for $\eta \leqslant 0$:
\[
  H(s) = \mathbb{L} [k(\eta)] = \int_0^{+\infty} k(\eta) \mathrm{e}^{-s\eta} d\eta, \ \ \
  H(\mathrm{i}\omega) = \mathbb{F} [k(\eta)] = \int_0^{+\infty} k(\eta) \mathrm{e}^{-\mathrm{i} \omega \eta} d\eta.
\]

The function $k(\eta)$ allows one to relate input and output signals in the integral form, namely
\begin{equation}\label{eqInOutIRF}
  x(t) = \int_0^t k(t-\tau) g(\tau) d\tau,
\end{equation}
and if the input signal $g(t)$ is the standard Gaussian white noise, then the output signal is represented as the stochastic integral over the standard Wiener process $w(t)$:
\begin{equation}\label{eqInOutIRFw}
  x(t) = \int_0^t k(t-\tau) dw(\tau).
\end{equation}

For multidimensional linear stationary systems, the transition matrix is similar to the function $k(\eta)$. The solution of a system of linear stochastic differential equations in the integral form based on the transition matrix is the well-known result~\cite{KloPla_92}.

Obviously, the relation~\eqref{eqInOutIRFw} defines the shaping filter. In this case, the output signal is correctly defined regardless of the value $m$, since for a linear stationary system the impulse response function is a continuous function that is representable in the most general form  as a linear combination of products of exponential functions, trigonometric functions (cosines and sines), and polynomials. According to the stochastic integral properties~\cite{Oks_03}, for any finite time $t$, the right-hand side of the relation~\eqref{eqInOutIRFw} is a random variable having a Gaussian distribution with zero mean and variance equal to the squared norm of the conjugate impulse response, i.e., the function $k(t-\tau)$ for a fixed time $t$, in the space $L_2([0,t])$:
\[
  \mathrm{E} x^2(t) = \int_0^t k^2(t-\tau) d\tau.
\]

For approximate modeling of the random process $x(t)$, it is sufficient to pass from the integral to the corresponding integral sum by the It\^o stochastic integral definition~\cite{Oks_03}. As in the description of the shaping filter by the stochastic differential equation or the system of stochastic differential equations with subsequent use of numerical methods for their solution, the numerical integration provides a discrete approximation of the random process $x(t)$.

Advantages of the considered approach: \nopagebreak

(1)\;the simple implementation of the numerical integration (there is no need to use the vector-valued random process just because of modeling one of its components or a linear combination of components);

(2)\;the generalization to linear non-stationary systems.

The disadvantage is associated with the need to find an analytical expression for the impulse response function.

For linear non-stationary systems, it is necessary to take into account that the impulse response function is a function of two variables $k(t,\tau)$, and it cannot be represented as $k(\eta)$. The shaping filter is then specified by a relation similar to~\eqref{eqInOutIRFw}. Moreover, this method can be used to describe shaping filters that cannot be defined by a stochastic differential equation of the form~\eqref{eqSDE}, even assuming that its coefficients depend on time. A typical example is the shaping filter for the fractional Brownian motion or the fractional Liouville process~\cite{Pic_SP11, Ryb_Comp25}.

\section{Description of the Shaping Filter Based on the Spectral Form}\label{secSpectral}

Next, we describe the spectral form of mathematical description for linear systems. For this, we should choose both the time interval $\mathds{T} = [0,T]$ and the orthonormal basis $\BasisT$ of the space $\LPT$. Then functions $\BasisTT$ form the orthonormal basis of the space $\LPTT$. Further, we can rewrite the relation~\eqref{eqInOutIRF} taking into account the physical realizability condition:
\[
  x(t) = \int_\mathds{T} k(t-\tau) g(\tau) d\tau,
\]
and we can represent the impulse response function as a functional series:
\begin{equation}\label{eqDefDNPF}
  k(t-\tau) = \sum\limits_{i,j=0}^\infty {W_{ij} q(i,t) q(j,\tau)}, \ \ \ W_{ij} = \int_{\mathds{T}^2} k(t-\tau) q(i,t) q(j,\tau) dt d\tau.
\end{equation}

Therefore,
\[
  x(t) = \int_\mathds{T} \sum\limits_{i,j=0}^\infty {W_{ij} q(i,t) q(j,\tau)} g(\tau) d\tau = \sum\limits_{i,j=0}^\infty W_{ij} \biggl[ \int_\mathds{T} q(j,\tau) g(\tau) d\tau \biggr] q(i,t),
\]
consequently, the expansion of the output signal into a functional series with respect to the orthonormal basis $\BasisT$ is obtained under the condition that the output signal belongs to $\LPT$. For this, it is sufficient but not necessary that both the input signal and the impulse response function belong to $\LPT$ and $\LPTT$, respectively.

Introducing notations for expansion coefficients $X_i$ and $G_i$ of signals $x(t)$ and $g(t)$:
\[
  X_i = \int_\mathds{T} q(i,t) x(t) dt, \ \ \ G_i = \int_\mathds{T} q(i,t) g(t) dt, \ \ \ i = 0,1,2,\dots,
\]
i.e.,
\[
  x(t) = \sum\limits_{i=0}^\infty X_i q(i,t), \ \ \ g(t) = \sum\limits_{i=0}^\infty G_i q(i,t),
\]
we obtain the following relation between them:
\[
  X_i = \sum\limits_{j=0}^\infty W_{ij} G_j,
\]
or in the matrix form
\begin{equation}\label{eqInOutSp}
  X = W G,
\end{equation}
where $X$ and $G$ are infinite column matrices with elements $X_i$ and $G_i$, respectively, $W$ is the infinite matrix with elements $W_{ij}$. According to~\cite{SolSemPeshNed_79}, $X$ and $G$ are non-stationary spectral characteristics of signals $x(t)$ and $g(t)$, and $W$ is the two-dimensional non-stationary transfer function of the linear system. The non-stationarity in~\cite{SolSemPeshNed_79} is understood in the sense that the right boundary $T$ of the interval $\mathds{T}$ can be a function of time. Moreover, the left boundary can also be a function of time in general, i.e., a linear system can be considered on the non-stationary time interval; however, non-stationary time intervals are not used in this paper. Further, we use the notation $\mathbb{S}$ for the spectral transform: $W = \mathbb{S}[k(t-\tau)]$, $X = \mathbb{S}[x(t)]$, and $G = \mathbb{S}[g(t)]$.

The spectral transform is similar in properties to the Laplace transform, in particular the relation~\eqref{eqInOutSp} for non-stationary spectral characteristics of input and output signals is similar to the relation
\[
  X(s) = H(s) G(s)
\]
in which $X(s)$ and $G(s)$ are the Laplace transforms of signals $x(t)$ and $g(t)$: $X(s) = \mathbb{L}[x(t)]$, $G(s) = \mathbb{L}[g(t)]$.

The classical Laplace transform is defined for exponential-type functions. However, a number of problems, including the description of linear systems, require the consideration of generalized functions such as the Dirac delta function and its derivatives. The Laplace transform is also defined for them. Similarly, the spectral transform domain is extended to generalized functions and their derivatives~\cite{SolSemPeshNed_79}. This allows one to describe elementary blocks of control systems such as proportional, integral, and derivative blocks in the spectral form (the impulse response function belongs to $\LPTT$ for the integral block only). An infinite matrix corresponds to each of them. For the proportional block with gain $k$, the two-dimensional non-stationary transfer function is equal to $kE$ (the impulse response function is $k(\eta) = k \delta(\eta)$, and the transfer function is $H(s) = k = \mathrm{const}$), where $E$ is the infinite identity matrix. For integral and derivative blocks, two-dimensional non-stationary transfer functions are denoted by $P^{-1}$ and $P$ (impulse response functions are $k(\eta) = 1(\eta)$ and $k(\eta) = \delta'(\eta)$, and transfer functions are $H(s) = 1/s$ and $H(s) = s$). In these relations, $1(\eta)$ and $\delta(\eta)$ are the unit step function and the Dirac delta function, respectively.

Elements $P_{ij}$ and $P_{ij}^{-1}$ of matrices $P$ and $P^{-1}$ are calculated as follows (matrices $P$ and $P^{-1}$ are mutually inverse):
\[
  P_{ij} = q(i,0) q(j,0) + \int_\mathds{T} q(i,t) \, \frac{dq(j,t)}{dt} \, dt, \ \ \ P_{ij}^{-1} = \int_\mathds{T} q(i,t) \int_0^t q(j,\tau) d\tau dt.
\]

The remarkable properties of the spectral transform $\mathbb{S}$ are that the two-dimensional non-stationary transfer function of a linear stationary system is expressed through the two-dimensional non-stationary transfer function of the derivative block (or additionally the two-dimensional non-stationary transfer function of the integral block), and coefficients of the left-hand and right-hand sides of the differential equation~\eqref{eqSDE}. And this expression is no more complicated than the expression of the transfer function~\eqref{eqHdef} corresponding to this equation (with remark that the transfer function is a function of a complex variable, and the two-dimensional non-stationary transfer function is the infinite matrix)~\cite{PanBor_16, SolSemPeshNed_79}:
\begin{equation}\label{eqDefDNPFbyODE1}
  \begin{aligned}
    W & = (a_n P^n + \ldots + a_1 P + a_0 E)^{-1} (b_m P^m + \ldots + b_1 P + b_0 E) \\
    & = P^{-n} (a_n E + \ldots + a_1 P^{-n+1} + a_0 P^{-n})^{-1} (b_m P^m + \ldots + b_1 P + b_0 E).
  \end{aligned}
\end{equation}

This representation is not unique. For example, the transfer function of the form
\[
  H(s) = \frac{b_m (s-\kappa_1)(s-\kappa_2) \cdots (s-\kappa_m)}{a_n (s-\lambda_1)(s-\lambda_2) \cdots (s-\lambda_n)},
\]
where $\kappa_1,\kappa_2,\dots,\kappa_m$ and $\lambda_1,\lambda_2,\dots,\lambda_n$ are the zeros of polynomials $M(s)$ and $D(s)$, respectively, implies the following representation of the two-dimensional non-stationary transfer function:
\begin{equation}\label{eqDefDNPFbyODE2}
  W = \frac{b_m}{a_n} \, (P - \lambda_1 E)^{-1}(P - \lambda_2 E)^{-1} \cdots (P - \lambda_n E)^{-1} (P - \kappa_1 E)(P - \kappa_2 E) \cdots (P - \kappa_m E).
\end{equation}

This means that the spectral transform table can be formed on the Laplace transform table, at least for impulse response functions of linear stationary systems~\cite{SolSemPeshNed_79}. The spectral form of mathematical description can also be used for linear non-stationary systems, but here the relation with the Laplace transform table is lost.

\begin{figure}[ht]
  \centering
  \includegraphics[scale = 1]{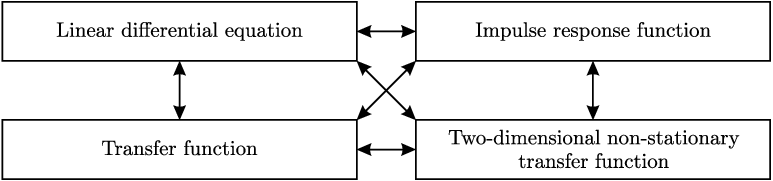}
  \vskip -1ex
  \caption{Various forms of mathematical description}\label{picForms}
\end{figure}

Thus, an arbitrary linear stationary system can be described in one of the four forms listed in Introduction, i.e., by the linear differential equation, impulse response function, transfer function, and two-dimensional non-stationary transfer function (for a linear non-stationary system, description by the transfer function is excluded). All forms of mathematical description are interconnected, and transitions from one form to another are possible (see Figure~\ref{picForms}). Some transitions are trivial, e.g., obtaining the transfer function or two-dimensional non-stationary transfer function from the differential equation. Other transitions seem more complex, e.g., the transition from the impulse response function to the differential equation.

Further, we give several examples of typical blocks of control systems, indicating the differential equation, impulse response function, transfer function, and two-dimensional non-stationary transfer function.

1. The aperiodic block (first-order system):
\begin{gather*}
  \theta x'(t) + x(t) = g(t), \ \ \ k(\eta) = \frac{1}{\theta} \, \mathrm{e}^{-\eta/\theta}, \\
  H(s) = \mathbb{L}[k(\eta)] = \frac{1}{\theta s+1}, \ \ \ W = \mathbb{S}[k(t-\tau)] = (\theta P + E)^{-1}.
\end{gather*}

2. The second-order aperiodic block (second-order system):
\begin{gather*}
  \theta^2 x''(t) + 2 \theta x'(t) + x(t) = g(t), \ \ \ k(\eta) = \frac{1}{\theta^2} \, \eta \, \mathrm{e}^{-\eta/\theta}, \\
  H(s) = \mathbb{L}[k(\eta)] = \frac{1}{(\theta s+1)^2}, \ \ \ W = \mathbb{S}[k(t-\tau)] = (\theta P + E)^{-2}.
\end{gather*}

3. The oscillatory block (second-order system):
\begin{gather*}
  \theta^2 x''(t) + 2 \xi \theta x'(t) + x(t) = g(t), \ \ \ k(\eta) = \frac{1}{\theta \sqrt{1-\xi^2}} \, \mathrm{e}^{-\xi \eta/\theta} \sin \frac{\sqrt{1-\xi^2} \, \eta}{\theta}, \\
  H(s) = \mathbb{L}[k(\eta)] = \frac{1}{\theta^2 s^2 + 2 \xi \theta s + 1}, \ \ \ W = \mathbb{S}[k(t-\tau)] = (\theta^2 P^2 + 2 \xi \theta P + E)^{-1}.
\end{gather*}

In the above expressions, $\theta > 0$ is the time constant, and $\xi \in (-1,1)$ is the damping coefficient.

To apply the spectral form of mathematical description for the shaping filter, it remains to represent the standard Gaussian white noise. We can use the same steps as at the beginning of this section but for the relation~\eqref{eqInOutIRFw}:
\[
  x(t) = \int_\mathds{T} \sum\limits_{i,j=0}^\infty {W_{ij} q(i,t) q(j,\tau)} dw(\tau) = \sum\limits_{i,j=0}^\infty W_{ij} \biggl[ \int_\mathds{T} q(j,\tau) dw(\tau) \biggr] q(i,t).
\]

According to the stochastic integral properties~\cite{Oks_03}, random variables
\[
  \cv_i = \int_\mathds{T} q(i,t) dw(t), \ \ \ i = 0,1,2,\dots,
\]
have a Gaussian distribution with zero mean and unit variance, since the integration is carried out over the centered Gaussian random process and all functions $\BasisT$ have unit norms. In addition, these random variables are pairwise uncorrelated, which follows from the orthogonality of functions $\BasisT$; therefore, they are independent. Thus, relations
\begin{equation}\label{eqInOutSpw}
  X = W \cv, \ \ \ x(t) = \sum\limits_{i=0}^\infty X_i q(i,t),
\end{equation}
where $\cv$ is the infinite random column matrix called the non-stationary spectral characteristic of the standard Gaussian white noise, define the shaping filter in the spectral form of mathematical description. Here, the white noise is interpreted as a random linear functional, and the definition of the non-stationary spectral characteristic of a linear functional~\cite{Ryb_Math23} is generalized. In addition, $X$ is the infinite random column matrix, which is also called the non-stationary spectral characteristic. There are no difficulties with the definition of the latter characteristic, since $x(t)$ is the random process with finite variance; its trajectories belong to $\LPT$. Moreover,
\begin{equation}\label{eqMSEnorm}
  \mathrm{E} \int_\mathds{T} x^2(t) dt = \int_\mathds{T} \int_0^t k^2(t-\tau) d\tau dt = \| k(t-\tau) \|_\LPTT^2.
\end{equation}

Relations~\eqref{eqInOutSpw} are the basis for approximate modeling of the random process $x(t)$. Such modeling assumes that the infinite matrix $W$ and the infinite random column matrices $X$ and $\cv$ are truncated to some finite size $L$ called the truncation order.

The representation of the impulse response function as a functional series and then the transition to the two-dimensional non-stationary transfer function are very important. There is a similar method based on canonical expansions of random processes~\cite{Sin_09}. Both the spectral form of mathematical description and the method of canonical expansions assume that the random process is represented as a functional series with random coefficients. However, for the canonical expansion, the linear transform is reduced to the modification of deterministic functions with the preservation of random coefficients, but in the spectral form, the linear transform preserves deterministic functions (orthonormal basis), and all modifications are carried out with random coefficients.

The considered approach provides a continuous (in continuous time) approximation of the random process $x(t)$ in contrast to approaches from previous sections.

Advantages of the spectral form of mathematical description for the shaping filter:

(1)\;the simple implementation (using algebraic operations with matrices: multiplication by a number, addition, multiplication, and inversion);

(2)\;approximate modeling of random processes in continuous time rather than discrete time;

(3)\;the generalization to linear non-stationary systems.

As an additional advantage, one can obtain the whitening filter (the output signal of the whitening filter is the standard Gaussian white noise~\cite{Ste_17}). For this, it is necessary to find the inverse matrix with respect to~$W$:
\[
  W^{-1} = (b_m P^m + \ldots + b_1 P + b_0 E)^{-1} (a_n P^n + \ldots + a_1 P + a_0 E).
\]

The disadvantage is the need to calculate elements of the two-dimensional non-stationary transfer function. However, this is compensated by the fact that for elementary blocks of control systems, necessary expressions are obtained for different bases formed by polynomials, trigonometric functions and related complex exponential functions, piecewise constant functions, as well as bases generated by some splines or wavelets. They form the library of algorithms for the spectral method~\cite{SolSemPeshNed_79, Rybin_03, Rybin_13}.

The two-dimensional non-stationary transfer function of the proportional block with variable gain has a more complex form compared to $kE$ (as for the proportional block with gain $k$); its elements should be calculated. But for linear non-stationary systems, the two-dimensional non-stationary transfer function is also a rational function with respect to the matrix $P$~\cite{SolSemPeshNed_79, PanBor_16}. The spectral form allows one to describe shaping filters for random processes such as the fractional Brownian motion or the fractional Liouville process~\cite{Ryb_Comp25}.

\section{Approximate Modeling of Random Processes and Error Analysis}\label{secExamples}

Following~\cite{KhruRum_AT24}, we consider three transfer functions that define the Dryden turbulence model in a generalized form, i.e., without specifying the constants and their physical meaning:
\[
  H_1(s) = \frac{\alpha}{\gamma s + 1}, \ \ \ H_2(s) = \frac{\alpha(\beta s + 1)}{(\delta s + 1)^2}, \ \ \ H_3(s) = \frac{\alpha s (\beta s + 1)}{(\gamma s + 1)(\delta s + 1)^2},
\]
where $\alpha,\beta,\gamma,\delta$ are non-zero parameters; $\beta,\gamma,\delta$ are pairwise different.

They correspond to the following two-dimensional non-stationary transfer functions according to the formulae~\eqref{eqDefDNPFbyODE1} and~\eqref{eqDefDNPFbyODE2}:
\begin{gather*}
  W_1 = \alpha (\gamma P + E)^{-1}, \ \ \ W_2 = \alpha (\delta P + E)^{-2} (\beta P + E), \\
  W_3 = \alpha (\gamma P + E)^{-1} (\delta P + E)^{-2} P (\beta P + E),
\end{gather*}
and shaping filters are specified by the relation~\eqref{eqInOutSpw}, i.e.,
\[
  X^l = W_l \cv^l, \ \ \ x_l(t) = \sum\limits_{i=0}^\infty X_i^l q(i,t), \ \ \ l = 1,2,3,
\]
where $X^l$ are non-stationary spectral characteristics of random processes $x_l(t)$: $X_l = \mathbb{S}[x_l(t)]$, and $\cv^l$ are non-stationary spectral characteristics of independent standard Gaussian white noises. Their elements are independent random variables $\cv_i^l$ having a Gaussian distribution with zero mean and unit variance.

In order for the above relations to be constructive, it is necessary to choose both the time interval $\mathds{T}$ and the orthonormal basis $\BasisT$ of $\LPT$. For approximate modeling of random processes $x_l(t)$, it is necessary to specify the truncation order $L$ and determine elements of square matrices $W_l$ of size $L \times L$ for this basis. Then
\[
  X^l = W_l \cv^l, \ \ \ X_i^l = \sum\limits_{j=0}^{L-1} W_{ij} \cv_j^l, \ \ \ i = 0,1,\dots,L-1; \ \ \ x_l(t) \approx \sum\limits_{i=0}^{L-1} X_i^l q(i,t), \ \ \ l = 1,2,3.
\]

Next, we describe how the mean square approximation error can be calculated. Let $\hat x(t)$ be the random process that is an approximation of the random process $x(t)$. The mean square approximation error is defined by the expression
\[
  \varepsilon = \mathrm{E} \int_\mathds{T} \bigl( x(t) - \hat x(t) \bigr)^2 dt.
\]

Random processes $x(t)$ and $\hat x(t)$ are output signals of shaping filters, which correspond to impulse response functions $k(\eta)$ and $\hat k(\eta)$, as well as two-dimensional non-stationary transfer functions $W$ and $\hat W$, i.e., ${\mathbb{S}[k(t-\tau)] = W}$ and $\mathbb{S}[\hat k(t-\tau)] = \hat W$. Based on the formula~\eqref{eqMSEnorm}, we obtain
\[
  \varepsilon = \| k(t-\tau) - \hat k(t-\tau) \|_\LPTT^2.
\]

Thus, the mean square approximation error is completely determined by the approximation error of the impulse response function in $\LPTT$. Here, it is appropriate to refer to works on the mean square approximation of iterated stochastic integrals~\cite{Kuz_DUPU23}. In particular, the mean square approximation error of the stochastic integral of the second multiplicity from the Milstein method is determined by the approximation error of the unit step function as a function of two variables, i.e., the impulse response function of the integral block.

The spectral transform is the orthogonal transform; it preserves the norm and inner product, i.e.,
\[
  \| k(t-\tau) - \hat k(t-\tau) \|_\LPTT^2 = \| W - \hat W \|^2,
\]
where $\| \cdot \|$ denotes the Euclidean norm of the matrix.

Further, we assume that the matrix $\tilde W$ is obtained from the two-dimensional non-stationary transfer function $W$ by the truncation, i.e., as a result of the transition from the infinite matrix to the matrix of size $L \times L$ ($\tilde W$ can be represented as the infinite matrix under the condition $\tilde W_{ij} = 0$ for $i \geqslant L$ or $j \geqslant L$). According to the Parseval's identity, we have
\[
  \| k(t-\tau) - \tilde k(t-\tau) \|_\LPTT^2 = \| k(t-\tau) \|_\LPTT^2 - \| \tilde W \|^2 = \int_\mathds{T} \int_0^t k^2(t-\tau) dt d\tau - \sum\limits_{i,j = 0}^{L-1} W_{ij}^2,
\]
where the impulse response function $\tilde k(\eta)$ corresponds to the two-dimensional non-stationary transfer function~$\tilde W$, i.e., $\mathbb{S}[\tilde k(t-\tau)] = \tilde W$.

When using the spectral form of mathematical description, the infinite matrices from the right-hand side of the formula~\eqref{eqDefDNPFbyODE1} are usually truncated, i.e., infinite matrices $P$ and $E$ are replaced by matrices $\tilde P$ and $\tilde E$ of size $L \times L$~\cite{SolSemPeshNed_79}. But then the result of calculations according to this formula will differ from $\tilde W$:
\[
  \hat W = (a_n \tilde P^n + \ldots + a_1 \tilde P + a_0 \tilde E)^{-1} (b_m \tilde P^m + \ldots + b_1 \tilde P + b_0 \tilde E) \neq \tilde W.
\]

A similar remark is hold for the formula~\eqref{eqDefDNPFbyODE2}:
\[
  \hat W = \frac{b_m}{a_n} \, (\tilde P - \lambda_1 \tilde E)^{-1}(\tilde P - \lambda_2 \tilde E)^{-1} \cdots (\tilde P - \lambda_n \tilde E)^{-1} (\tilde P - \kappa_1 \tilde E)(\tilde P - \kappa_2 \tilde E) \cdots (\tilde P - \kappa_m \tilde E) \neq \tilde W,
\]
consequently, the additional error arises, namely
\[
  \| \tilde k(t-\tau) - \hat k(t-\tau) \|_\LPTT^2 = \| \tilde W - \hat W \|^2 = \sum\limits_{i,j = 0}^{L-1} (W_{ij} - \hat W_{ij})^2.
\]

Therefore,
\begin{equation}\label{eqError}
  \begin{aligned}
    \varepsilon & = \| k(t-\tau) - \tilde k(t-\tau) + \tilde k(t-\tau) - \hat k(t-\tau) \|_\LPTT^2 \\
    & = \| k(t-\tau) - \tilde k(t-\tau) \|_\LPTT^2 + \| \tilde k(t-\tau) - \hat k(t-\tau) \|_\LPTT^2 \\
    & = \underbrace{\| k(t-\tau) \|_\LPTT^2 - \| \tilde W \|^2}_{\varepsilon_1} + \underbrace{\| \tilde W - \hat W \|^2 \vphantom{\|_\LPTT^2}}_{\varepsilon_2},
  \end{aligned}
\end{equation}
where the second equality follows from the orthogonality of functions $k(t-\tau) - \tilde k(t-\tau)$ and $\tilde k(t-\tau) - \hat k(t-\tau)$, since they are represented by different subsets of functions $\BasisTT$ as the functional series. The errors $\varepsilon$, $\varepsilon_1$, and $\varepsilon_2$ depend on the truncation order $L$.

Any of the three transfer functions given above can be represented as a sum of fractions:
\[
  H(s) = \frac{\Delta_1}{\gamma s + 1} + \frac{\Delta_2}{\delta s + 1} + \frac{\Delta_3}{(\delta s + 1)^2},
\]
where coefficients $\Delta_1,\Delta_2,\Delta_3$  are uniquely determined by parameters $\alpha,\beta,\gamma,\delta$ for each transfer function separately. We do not give the general formulae for coefficients but point out that $\Delta_2 = \Delta_3 = 0$ for $H_1(s)$ and $\Delta_1 = 0$ for $H_2(s)$. Such a sum of fractions corresponds to the impulse response function
\[
  k(\eta) = \frac{\Delta_1}{\gamma} \, \mathrm{e}^{-\eta/\gamma} + \frac{\Delta_2}{\delta} \, \mathrm{e}^{-\eta/\delta} + \frac{\Delta_3}{\delta^2} \, \eta \, \mathrm{e}^{-\eta/\delta}.
\]

\begin{figure}[ht]
  \centering
  \includegraphics[scale = 1]{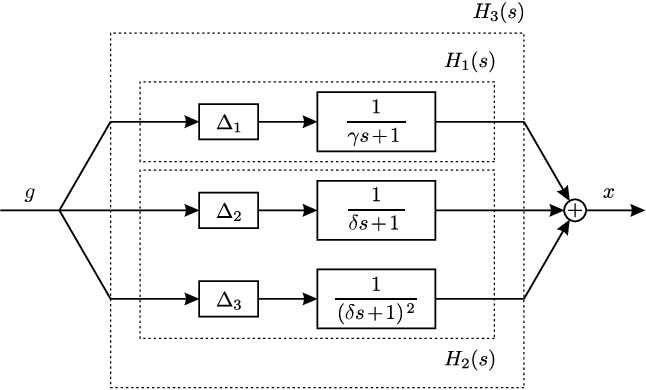}
  \vskip -1ex
  \caption{The shaping filter for the Dryden turbulence model}\label{picDiagram}
\end{figure}

This implies that the shaping filter for the random process $x_1(t)$ is specified by the aperiodic block; for $x_2(t)$ it is described by a parallel connection of the aperiodic block and the second-order aperiodic block; for $x_3(t)$ it is described by a parallel connection of three blocks: two aperiodic blocks and the second-order aperiodic block (see Figure~\ref{picDiagram}). The time constants of aperiodic blocks are $\gamma$ and $\delta$. Such an interpretation allows one to obtain systems of stochastic differential equations defining random processes $x_2(t)$ and $x_3(t)$, which differ from the result of applying the methods from~\cite{KhruRum_AT24, PugSin_90}.

As the orthonormal basis $\BasisT$ of $\LPT$, we choose cosines, which provide high accuracy of the approximation in problems where approximate modeling of the Wiener process is required~\cite{Ryb_Math23}. The definition of this basis and relations to calculate elements of two-dimensional non-stationary transfer functions of integral and derivative blocks, as well as typical blocks such as aperiodic block, second-order aperiodic block, and oscillatory block are given in Appendix.

Table~\ref{tabErrors1} shows the results of calculations by the formula~\eqref{eqError} depending on the truncation order $L$ under the condition $T = 5$, i.e., $\mathds{T} = [0,5]$, and with the following parameters: $\alpha = 1$, $\beta = 2$, $\gamma = 3$, $\delta = 4$. Additionally, we indicate non-zero coefficients $\Delta_l$ and squared norms of impulse response functions $k_l(t-\tau)$ corresponding to the given transfer functions, $l = 1,2,3$:
\begin{align*}
  & H_1(s) \colon & & \Delta_1 = 1, & & \hskip -25em \| k_1(t-\tau) \|_\LPTT^2 = \frac{1}{4} \, \mathrm{e}^{-10/3} + \frac{7}{12} \approx 0.592252; \\
  & H_2(s) \colon & & \Delta_2 = \Delta_3 = 1/2, & & \hskip -25em \| k_2(t-\tau) \|_\LPTT^2 = \frac{177}{256} \, \mathrm{e}^{-5/2} + \frac{7}{64} \approx 0.166129; \\
  & H_3(s) \colon & & \Delta_1 = -1, \ \ \ \Delta_2 = 3/2, \ \ \ \Delta_3 = -1/2, \vphantom{\frac{1}{2}}\\
  & & & \ \ \ \| k_3(t-\tau) \|_\LPTT^2 = \frac{17}{256} \, \mathrm{e}^{-5/2} + \frac{1}{4} \, \mathrm{e}^{-10/3} - \frac{75}{343} \, \mathrm{e}^{-35/12} + \frac{379}{65856} \approx 0.008292.
\end{align*}

This table contains values $\varepsilon$ and also, for comparison, values $\varepsilon_1$ (in brackets). From these results, one can evaluate the convergence rate: $\varepsilon \approx C/L$, $\varepsilon_1 \approx C_1/L$, where $C,C_1 > 0$ are constants depending on the transfer function, i.e., if we double the truncation order, then we approximately halve the mean square approximation error.

\begin{table}[ht]
\begin{center}
\caption{The approximation errors $\varepsilon$ ($\varepsilon_1$)}\label{tabErrors1}
\begin{tabular}{|c|c|c|c|c|c|c|c|}
  \hline
  & $L = 4$ & $L = 8$ & $L = 16$ & $L = 32$ & $L = 64$ & $L = 128$ & $L = 256$ \\
  \hline
  \hline
  $x_1(t)$ & 0.125603 & 0.055877 & 0.024617 & 0.011053 & 0.005098 & 0.002411 & 0.001162 \\
  & (0.091720) & (0.041186) & (0.019233) & (0.009241) & (0.004518) & (0.002231) & (0.001108) \\
  \hline
  $x_2(t)$ & 0.022098 & 0.008628 & 0.003592 & 0.001576 & 0.000720 & 0.000340 & 0.000164 \\
  & (0.013487) & (0.005851) & (0.002711) & (0.001300) & (0.000635) & (0.000314) & (0.000156) \\
  \hline
  $x_3(t)$ & 0.001981 & 0.000877 & 0.000385 & 0.000173 & 0.000080 & 0.000038 & 0.000018 \\
  & (0.001451) & (0.000645) & (0.000301) & (0.000144) & (0.000071) & (0.000035) & (0.000017) \\
  \hline
\end{tabular}
\end{center}
\end{table}

For small $L$, there is a significant relative error. However, the analysis of the results obtained for approximate modeling of random processes that are described by first-order stochastic differential equations, using both the spectral form of mathematical description and the Euler--Maruyama method, shows that $L$ should be comparable with the number of steps in the numerical scheme~\cite{Ryb_DUPU20}. Therefore, it is inappropriate to choose small truncation orders.

For large $L$, the good approximation accuracy is achieved. So, the truncation order can be chosen to achieve the desired accuracy.

The additional error $\varepsilon_2$ in the formula~\eqref{eqError} can be reduced to zero (then $\varepsilon = \varepsilon_1$) if we will not use the formulae~\eqref{eqDefDNPFbyODE1} and~\eqref{eqDefDNPFbyODE2} with matrices $\tilde P$ and $\tilde E$ of size $L \times L$. In the example under consideration, this is achieved by using the following general expression for the two-dimensional non-stationary transfer function:
\[
  W = \mathbb{S}[k(t-\tau)] = \Delta_1 A_\gamma + \Delta_2 A_\delta + \Delta_3 A_\delta^2,
\]
where $A_\gamma$ and $A_\delta$ are two-dimensional non-stationary transfer functions of aperiodic blocks, $A_\delta^2$ is the two-dimensional non-stationary transfer function of the second-order aperiodic block, and the subscript indicates the time constant of the aperiodic block. Appendix contains all necessary relations to calculate their elements according to the formula~\eqref{eqDefDNPF}. Appendix also give relations for elements of the two-dimensional non-stationary transfer function of the oscillatory block.

On the one hand, the derivation of such relations is quite complex. On the other hand, they can be used not only in the considered problem of synthesizing a shaping filter but also in other problems related to the analysis, synthesis, and identification of linear continuous systems. Such relations complement the library of algorithms for the spectral method.

\begin{figure}[ht]
  \centering
  \includegraphics[scale = 1]{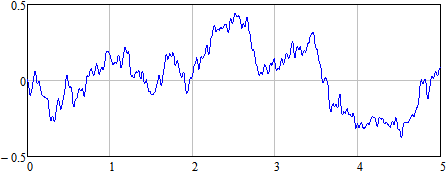}
  \vskip -2ex
  \caption{Trajectory of the random process $x_1(t)$ with power spectral density $S_1(\omega) = |H_1(\mathrm{i} \omega)|^2$}\label{picPath1}
  \vskip 2ex
  \includegraphics[scale = 1]{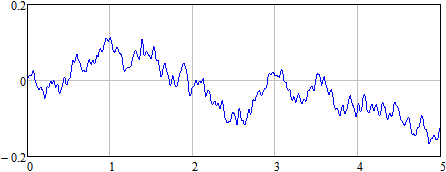}
  \vskip -2ex
  \caption{Trajectory of the random process $x_2(t)$ with power spectral density $S_2(\omega) = |H_2(\mathrm{i} \omega)|^2$}\label{picPath2}
  \vskip 2ex
  \includegraphics[scale = 1]{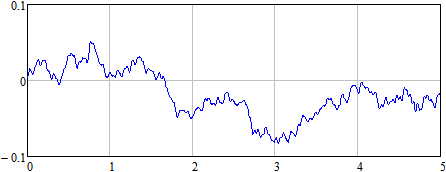}
  \vskip -2ex
  \caption{Trajectory of the random process $x_3(t)$ with power spectral density $S_3(\omega) = |H_3(\mathrm{i} \omega)|^2$}\label{picPath3}
\end{figure}

Sample trajectories of random processes $x_l(t)$ for the truncation order $L = 256$, $l = 1,2,3$, are shown in Figures~\ref{picPath1}--\ref{picPath3}.

Note that for approximate modeling of random processes corresponding to the von Karman turbulence model (it is recommended by the standard~\cite{MIL-STD-1797A} along with the Dryden turbulence model), the spectral form of mathematical description can also be used. Corresponding transfer functions are such that shaping filters are specified by a parallel connection of two, three, or four aperiodic blocks.

In addition, we consider the example of the shaping filter defined by the following transfer function:
\[
  H_4(s) = \frac{1}{4s^2 + 2s + 1},
\]
i.e., it corresponds to the oscillatory block with the time constant $\theta = 2$ and the damping coefficient $\xi = 1/2$.

In this case, the two-dimensional non-stationary transfer function has the form
\begin{gather*}
  W_4 = (4P^2 + 2P + E)^{-1} = K_{2,\,1/2},
\end{gather*}
where $K_{2,\,1/2}$ corresponds to the notation of the two-dimensional non-stationary transfer function $K_{\theta,\xi}$ of the oscillatory block, and the shaping filter is determined as above:
\[
  X^4 = W_4 \cv^4, \ \ \ x_4(t) = \sum\limits_{i=0}^\infty X_i^4 q(i,t),
\]
where $X^4$ and $\cv^4$ are non-stationary spectral characteristics of the random process $x_4(t)$ and the standard Gaussian white noise, respectively.

For approximate modeling of the random process $x_4(t)$, we have
\[
  X^4 = W_4 \cv^4, \ \ \ X_i^4 = \sum\limits_{j=0}^{L-1} W_{ij} \cv_j^4, \ \ \ i = 0,1,\dots,L-1; \ \ \ x_4(t) \approx \sum\limits_{i=0}^{L-1} X_i^4 q(i,t).
\]

In addition, we present the impulse response function and its squared norm:
\begin{gather*}
  k_4(\eta) = \frac{1}{\sqrt{3}} \, \mathrm{e}^{-t/4} \sin \frac{\sqrt{3} t}{4}, \\
  \| k_4(t-\tau) \|_\LPTT^2 = \biggl( \frac{2}{3} + \frac{1}{12} \cos \frac{5 \sqrt{3}}{2} + \frac{\sqrt{3}}{12} \sin \frac{5 \sqrt{3}}{2} \biggr) \mathrm{e}^{-5/4} + \frac{1}{2} \approx 0.541179,
\end{gather*}
since this is necessary to calculate the mean square approximation error exactly.

The results of calculations by the formula~\eqref{eqError} for this example are given in Table~\ref{tabErrors2} under the same condition $T = 5$; the orthonormal basis is without changes. These results show a higher convergence rate: $\varepsilon \approx C/L^3$, $\varepsilon_1 \approx C_1/L^3$, where $C,C_1 > 0$, i.e., if we double the truncation order, then we approximately decrease the mean square approximation error by about eight times. But in contrast to the results from Table~\ref{tabErrors1}, here the approximation errors $\varepsilon$ and $\varepsilon_1$ differ by about ten times.

\begin{table}[ht]
\begin{center}
\renewcommand{\arraystretch}{1.1}
\caption{The approximation errors $\varepsilon$ ($\varepsilon_1$)}\label{tabErrors2}
\begin{tabular}{|c|c|c|c|c|c|c|c|}
  \hline
  & $L = 4$ & $L = 8$ & $L = 16$ & $L = 32$ & $L = 64$ & $L = 128$ & $L = 256$ \\
  \hline
  \hline
  $x_4(t)$ & 0.035217 & 0.004319 & 0.000524 & 0.000065 & $8.21 \cdot 10^{-6}$ & $1.06 \cdot 10^{-6}$ & $1.39 \cdot 10^{-7}$ \\
  & (0.005314) & (0.000531) & (0.000060) & ($7.14 \cdot 10^{-6}$) & ($8.71 \cdot 10^{-7}$) & ($1.08 \cdot 10^{-7}$) & ($1.34 \cdot 10^{-8}$) \\
  \hline
\end{tabular}
\end{center}
\end{table}

\begin{figure}[ht]
  \centering
  \includegraphics[scale = 1]{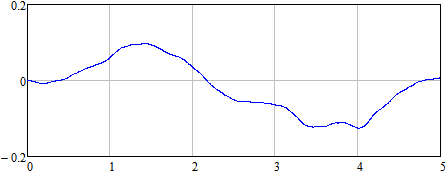}
  \vskip -2ex
  \caption{Trajectory of the random process $x_4(t)$ with power spectral density $S_4(\omega) = |H_4(\mathrm{i} \omega)|^2$}\label{picPath4}
\end{figure}

Figure~\ref{picPath4} depicts a sample trajectory of the random process $x_4(t)$ for the truncation order $L = 256$.

\section{Conclusions}\label{secConcl}

The main result of the paper is a new approach to synthesizing a shaping filter based on the spectral form of mathematical description. The considered approach allows one to obtain an approximation of a random process in continuous time and to calculate the mean square approximation error exactly. It is not presented separately but in relation to different forms of mathematical description for linear continuous systems. Along with the theoretical explanation, the paper includes all relations required to implement a shaping filter.

\section*{Appendix}

This part contains all necessary relations to calculate elements of two-dimensional non-stationary transfer functions of some elementary and typical blocks of control systems. As the orthonormal basis of the space space $\LPT$, $\mathds{T} = [0,T]$, cosines are used:
\[
  q(i,t) = \left\{ \begin{array}{ll}
    \displaystyle \sqrt{\frac{1}{T}} & \text{for} ~ i = 0 \\ [-2.0ex] \\
    \displaystyle \sqrt{\frac{2}{T}} \, \cos \frac{i \pi t}{T} & \text{for} ~ i = 1,2,\dots
  \end{array} \right.
\]

1.\;Elements of the two-dimensional non-stationary transfer function $P^{-1}$ of the integral block: $P_{ij}^{-1} = T c_{ij}$, where
\begin{gather*}
  c_{00} = \frac{1}{2}, \ \ \ c_{0i} = -c_{i0} = \frac{\sqrt{2} [1-(-1)^i]}{i^2 \pi^2}, \ \ \ c_{ii} = 0, \ \ \ c_{ij} = -c_{ji} = \frac{2[(-1)^{i+j} - 1]}{(i^2-j^2) \pi^2}, \\
  i = 1,2,3,\dots, \ \ \ j = 1,2,\dots,i-1.
\end{gather*}

2.\;Elements of the two-dimensional non-stationary transfer function $P$ of the derivative block: $P_{ij} = c_{ij} / T$, where
\begin{gather*}
  c_{00} = 1, \ \ \ c_{0i} = (-1)^i c_{i0} = (-1)^i \sqrt{2}, \ \ \ c_{ii} = 2, \ \ \ c_{ij} = (-1)^{i+j} c_{ji} = \frac{2[i^2 - (-1)^{i+j} j^2]}{i^2 - j^2}, \\
  i = 1,2,3,\dots, \ \ \ j = 1,2,\dots,i-1.
\end{gather*}

For the following two items, we assume that
\[
  \theta > 0, \ \ \ \mu = -\frac{1}{\theta}, \ \ \ \phi_i^\pm = \mu^2 T^2 \pm i^2 \pi^2.
\]

3.\;Elements of the two-dimensional non-stationary transfer function $A_\theta$ of the aperiodic block: ${(A_\theta)_{ij} = -\mu c_{ij}}$, where
\begin{align*}
  c_{00} & = \frac{\mathrm{e}^{\mu T} - \mu T - 1}{\mu^2 T}, \ \ \ c_{0i} = (-1)^i c_{i0} = \sqrt{2} T \, \frac{\mathrm{e}^{\mu T} - (-1)^i}{\phi_i^+}, \\
  c_{ij} & = (-1)^{i+j} c_{ji} = \frac{2T}{\phi_j^+} \biggl( \mu^2 T^2 \, \frac{(-1)^i \mathrm{e}^{\mu T} - 1}{\phi_i^+} - \gamma_{ij} \biggr), \\
  \gamma_{ij} & = \left\{
    \begin{array}{ll}
      \displaystyle \frac{\mu T}{2} & \text{for} ~ i = j \\ [-2.0ex] \\
      \displaystyle \frac{j^2[1 - (-1)^{i+j}]}{i^2-j^2} & \text{for} ~ i \neq j,
    \end{array}
  \right. \ \ \
  i = 1,2,3,\dots, \ \ \ j = 1,2,\dots,i.
\end{align*}

4.\;Elements of the two-dimensional non-stationary transfer function $A_\theta^2$ of the second-order aperiodic block: $(A_\theta^2)_{ij} = \mu^2 c_{ij}$, where
\begin{align*}
  c_{00} & = \frac{(\mu T - 2) \, \mathrm{e}^{\mu T} + \mu T + 2}{\mu^3 T}, \ \ \ c_{0i} = (-1)^i c_{i0} = \sqrt{2} T^2 \, \frac{2 \mu T [(-1)^i - \mathrm{e}^{\mu T}] + \phi_i^+ \mathrm{e}^{\mu T}}{(\phi_i^+)^2}, \\
  c_{ij} & = (-1)^{i+j} c_{ji} = \frac{2 \mu T^3}{\phi_i^+ \phi_j^+} \biggl( [1 - (-1)^i \mathrm{e}^{\mu T}] \biggl[ \frac{\phi_i^-}{\phi_i^+} + \frac{\phi_j^-}{\phi_j^+} \biggr] + (-1)^i \mu T \mathrm{e}^{\mu T} \biggr) + \frac{\zeta_{ij} T^2}{(\phi_j^+)^2}, \\
  \zeta_{ij} & = \left\{
    \begin{array}{ll}
      \displaystyle \phi_j^- & \text{for} ~ i = j \\ [-2.0ex] \\
      \displaystyle \frac{4 \mu j^2 T [1 - (-1)^{i+j}]}{i^2-j^2} & \text{for} ~ i \neq j,
    \end{array}
  \right. \ \ \
  i = 1,2,3,\dots, \ \ \ j = 1,2,\dots,i.
\end{align*}

For the following item, we assume that
\begin{gather*}
  \theta > 0, \ \ \ \xi \in (-1,1), \ \ \ \mu = -\frac{\xi}{\theta}, \ \ \ \nu = \frac{\sqrt{1-\xi^2}}{\theta}, \\
  \lambda = \sqrt{\mu^2 + \nu^2} = \frac{1}{\theta}, \ \ \ \eta = \sqrt{\mu^2 - \nu^2}, \ \ \ \psi_i^\pm = \lambda^2 T^2 \pm i^2 \pi^2.
\end{gather*}

5.\;Elements of the two-dimensional non-stationary transfer function $K_{\theta,\xi}$ of the oscillatory block: $(K_{\theta,\xi})_{ij} = c_{ij} / (\theta \sqrt{1-\xi^2})$, where
\begin{align*}
  c_{00} & = \frac{2 \mu \nu (1 - \mathrm{e}^{\mu T} \cos \nu T) + \eta^2 \mathrm{e}^{\mu T} \sin \nu T + \nu \lambda^2 T}{\lambda^4 T}, \\
  c_{0i} & = (-1)^i c_{i0} = \sqrt{2} T \, \frac{2 \mu \nu T^2 [(-1)^i - \mathrm{e}^{\mu T} \cos \nu T] + (\eta^2 T^2 + i^2 \pi^2) \, \mathrm{e}^{\mu T} \sin \nu T}{(\psi_i^+)^2 - (2 \pi \nu i T)^2}, \\
  c_{ij} & = (-1)^{i+j} c_{ji} = 2 T^3 \biggl( \frac{2 \mu \nu (\lambda^4 T^4 - i^2 j^2 \pi^4) [1 - (-1)^i \mathrm{e}^{\mu T} \cos \nu T]}{[(\psi_i^+)^2 - (2 \pi \nu i T)^2][(\psi_j^+)^2 - (2 \pi \nu j T)^2]} \\
  & \ \ \ {} + \frac{(-1)^i (\eta^2 \lambda^4 T^4 + \pi^2 [\lambda^4 (i^2 + j^2) T^2 + (\pi i j \eta)^2]) \, \mathrm{e}^{\mu T} \sin \nu T}{[(\psi_i^+)^2 - (2 \pi \nu i T)^2][(\psi_j^+)^2 - (2 \pi \nu j T)^2]} \biggr) + \frac{\nu \kappa_{ij} T^2}{(\psi_j^+)^2 - (2 \pi \nu j T)^2}, \\
  \kappa_{ij} & = \left\{
    \begin{array}{ll}
      \displaystyle \psi_j^- & \text{for} ~ i = j \\ [-2.0ex] \\
      \displaystyle \frac{4 \mu j^2 T [1 - (-1)^{i+j}]}{i^2-j^2} & \text{for} ~ i \neq j,
    \end{array}
  \right. \ \ \
  i = 1,2,3,\dots, \ \ \ j = 1,2,\dots,i.
\end{align*}

Relations from items 1 and 2 are given in~\cite{SolSemPeshNed_79, PanBor_16}, relations from item 3 are obtained in~\cite{Ryb_IJMSSC20}, and relations from items 4 and 5 are published for the first time.

\end{document}